\documentclass[english,aps,pre, reprint, twocolumn]{revtex4-1}
\usepackage[T1]{fontenc}
\usepackage[latin9]{inputenc}
\usepackage{float}
\usepackage{amsmath}
\usepackage{graphicx}
\usepackage[caption=false]{subfig}
\makeatletter
\@ifundefined{textcolor}{}
{%
 \definecolor{BLACK}{gray}{0}
 \definecolor{WHITE}{gray}{1}
 \definecolor{RED}{rgb}{1,0,0}
 \definecolor{GREEN}{rgb}{0,1,0}
 \definecolor{BLUE}{rgb}{0,0,1}
 \definecolor{CYAN}{cmyk}{1,0,0,0}
 \definecolor{MAGENTA}{cmyk}{0,1,0,0}
 \definecolor{YELLOW}{cmyk}{0,0,1,0}
}

\makeatother

\usepackage{babel}
\begin{document}

\title{Rapid phase calibration and increased efficiency of a spatial light modulator using novel phase masks and an iterative algorithm}

\author{Amar Deo Chandra}

\author{Ayan Banerjee}
\email{ayan@iiserkol.ac.in}
\affiliation{Center of Excellence in Space Sciences India, Indian Institute of Science Education and Research Kolkata, Mohanpur 741246, West Bengal, India}
\affiliation{Dept of Physical Sciences, Indian Institute of Science Education
and Research Kolkata, Mohanpur, Kolkata, India 741235}


\begin{abstract} 
We develop an improved phase calibration method of a reflective spatial light modulator (SLM) using interferometry by employing novel phase masks. We generate the optimised phase masks by using Iterative Fourier Transform Algorithm (IFTA) and demonstrate that they perform with 18$\%$ better efficiency over global linear corrections in the look-up table (LUT). In addition, we also implement global linear corrections in the LUT and correspondingly modify the phase encoding of blazed gratings to improve the efficiency of our phase-limited SLM. In the process, we definitively determine the actual maximum phase throw of our SLM which provides a recipe for users to verify supplier specifications.Besides obtaining array of 1D/2D spots having high uniformity (~90$\%$) using IFTA, our result exemplifies the use of iterative algorithms for improving efficiency of phase limited SLMs. Finally, our method enables threefold faster phase measurements, and to the best of our knowledge, is the first endeavour directed towards enabling rapid phase characterisation of an SLM using interferometric measurements and can have very useful applications in settings which often require faster phase calibrations. 
\end{abstract}
\maketitle
\section{Introduction}
Spatial Light Modulators (SLMs) are dynamic optical elements which can be used to modulate the amplitude or the phase of an incident light beam. They have given rise to a plethora of applications by leveraging their modulation characteristics in the field of beam shaping \cite{grier2003revolution}, optical trapping \cite{grier2003revolution, padgett2011holographic,dholakia2011shaping}, super-resolution imaging \cite{ochoa2013super} and wavefront-correction \cite{love1997wave,wulff2006aberration}. Applications such as beam shaping and optical trapping entail static modification of the phase distribution of a light beam falling on the SLM. On the other hand, applications such as wavefront-correction require dynamic changes in the phase distribution of the incident light beam. It is necessary to measure the phase response of the SLM in ambient laboratory conditions because the phase response of an SLM varies with parameters such as incident laser power \cite{engstrom2013calibration}, operating temperature \cite{zhang2016temperature} and wavelength of incident light \cite{lizana2008wavelength}. Thus, use of SLMs in any phase modulation settings with high fidelity necessitates precise knowledge of the phase response of the device. The efficacy of SLMs in many applications such as beam shaping and holographic optical trapping depend on the diffraction efficiency of the device which is dependent on criteria such as the fill factor \cite{arrizon1999implementation}, pixel size, input polarisation, maximum phase modulation and non-linear phase response of the SLM \cite{moreno2004modulation, bowman2011optimisation}. There are a variety of SLMs currently available in the market having phase modulation depths ranging anywhere from about $\pi$ radians to about $2\pi$ radians and higher. Phase limited SLMs ($\phi_{max} < 2\pi$) are economical compared to the high-end ones but exhibit limited diffraction efficiency. However, recent studies \cite{moreno2004modulation, bowman2011optimisation} have shown that it is possible to improve the efficiency of phase limited SLMs by changing their phase response by modifying the global look-up table (LUT) of these devices. Applications such as beam shaping and optical trapping often require splitting the input beam into an array of spots. An interesting study \cite{curtis2005symmetry} has demonstrated that non-iterative algorithms perform poorly in efficiency compared to iterative algorithms such as the Gerchberg-Saxton (GS) algorithm \cite{gs1972algo} and other modified iterative algortihms \cite{fienup1980iterative,haist1997computer,sinclair2004assembly,ripoll2004review} while generating an array of symmetric spots. 

There are numerous methods discussed in literature such as intensity correlation \cite{yun2005interframe,yue2012characterization} and interferometry \cite{bergeron1995phase,reichelt2013spatially, villalobos2015phase, fuentes2016interferometric} to characterize the phase response of an SLM. Interferometric methods are accurate but are prone to vibrations and local refractive index variations. To alleviate this problem, a workaround employing binary phase masks having variable and fixed reference gray levels has been used for phase calibration of an SLM. This conventional method is time-consuming and cumbersome and is less suitable for applications which often require faster phase calibration. Iterative algorithms certainly have an edge over the non-iterative ones for generating symmetric patterns but high computational costs inhibit their  uses in real-time applications. This problem has been alleviated to some extent due to the advent of faster computers and algorithms programmed on Graphics Processing Unit (GPU) \cite{bianchi2010real}. There is a pressing need to implement robust iterative algorithms which have rapid convergence to reduce computational requirements for real-time applications at limited computational budget. In addition, the efficacy of iterative algorithms are measured by fast convergence, efficiency and the uniformity of the optically reconstructed pattern. It is a challenging task to design iterative algorithms which simultaneously perform well on all these three criteria. The primary causes of limited diffraction efficiency in phase-limited SLMs are reduced phase throw and non-linear phase modulation \cite{moreno2004modulation, bowman2011optimisation}. To mitigate this problem a solution using global linear corrections in the look-up table (LUT) of the SLM has been suggested \cite{bowman2011optimisation} which reduces the mismatch between the desired phase and the actual phase exhibited by the SLM. However, these mitigations are only global linear corrections to the LUT, and approaches which can better approximate the non-linear phase response of the SLM can further improve the efficiency of SLMs.

We report an improved method of phase calibration of a reflective SLM using novel phase masks which allows measurement of simultaneous phase shifts of three gray levels at a time with respect to the reference gray level. As a result, our method enables threefold faster phase measurements over conventional methods without modifying the experimental setup. In addition, our method can easily be implemented in any phase measurement settings. To our knowledge, our improved method is the first attempt towards enabling rapid phase characterization of an SLM and can have a large number of applications in settings which often demand faster phase calibration. Most importantly, it also exemplifies simultaneous differential tuning of an interferometer which can have immense advantages in the context of real-time, multi-wavelength spectroscopic applications. This can enable collection of simultaneous spectral information at four different wavelengths using a single interferometer to better investigate phenomena which evolve on timescales smaller than the tuning interval of the interferometer. In our method, we generate optimized phase masks using an iterative algorithm \cite{gu2012generation} and improve the efficiency by about 18$\%$ over global linear corrections in the LUT. Our result illustrates usage of iterative algorithms to improve the efficiency of phase limited SLMs which is a less explored area of research. We obtain rapid convergence of optimised phase solutions (within about 10 iterations for 1x5 array of spots) and achieve high uniformity (about 90$\%$) while generating an array of spots in 1D/2D which can have very useful applications in avenues such as holographic optical trapping and near real-time adaptive wavefront sensing. However, we note that this iterative algorithm gives moderate efficiency (about $44\%$) which is due to the reduced phase throw of our SLM. We corroborate this observation from the phase calibration curve of our SLM which shows limited phase depth of about $0.8\pi$. Thus, this is actually a useful means to check for actual performance of an SLM against manufacturer specifications. 

\section{Methods}
\begin{figure}[ht!]
\centering\includegraphics[width=\linewidth]{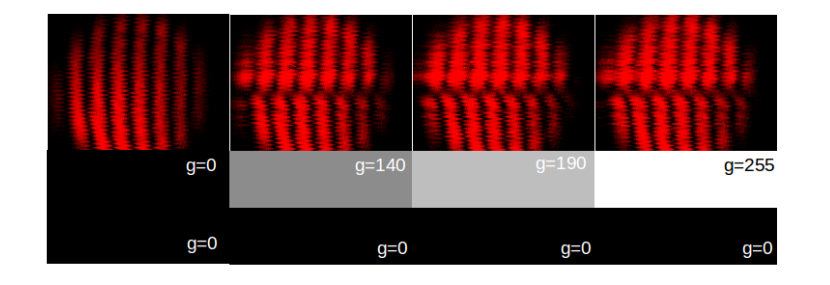}
\caption{Interferograms obtained using different binary phase masks addressed on the SLM. The gray levels used in each binary phase mask are indicated in the figure using the notation g=x where x indicates the gray level value.}
\label{Fig. 1}
\end{figure}

We use a Michelson interferometer for phase characterization of the reflective SLM (LC-R 1080) from HOLOEYE Photonics AG . The mirror in one of the arms of the interferometer is replaced by the reflective SLM. We study and characterise the phase response of our SLM at 671 nm. Since phase measurements are accurate but are sensitive to vibrations and local turbulence, we generate binary phase masks which are horizontally divided into variable gray level and fixed reference gray level in order to alleviate this problem. Our SLM permits 8 bit gray level addressing and we vary grayvalues from 0-255 in steps of 10 gray level (15 for the last step) and record twenty interferograms for each step. Fig. \ref{Fig. 1} shows the montage of interferograms obtained using different binary phase masks displayed on the SLM. This exemplifies that differential phase shifts are produced on displaying different binary gray level masks on the SLM.

\begin{figure}[ht!]
\centering\includegraphics[width=\linewidth]{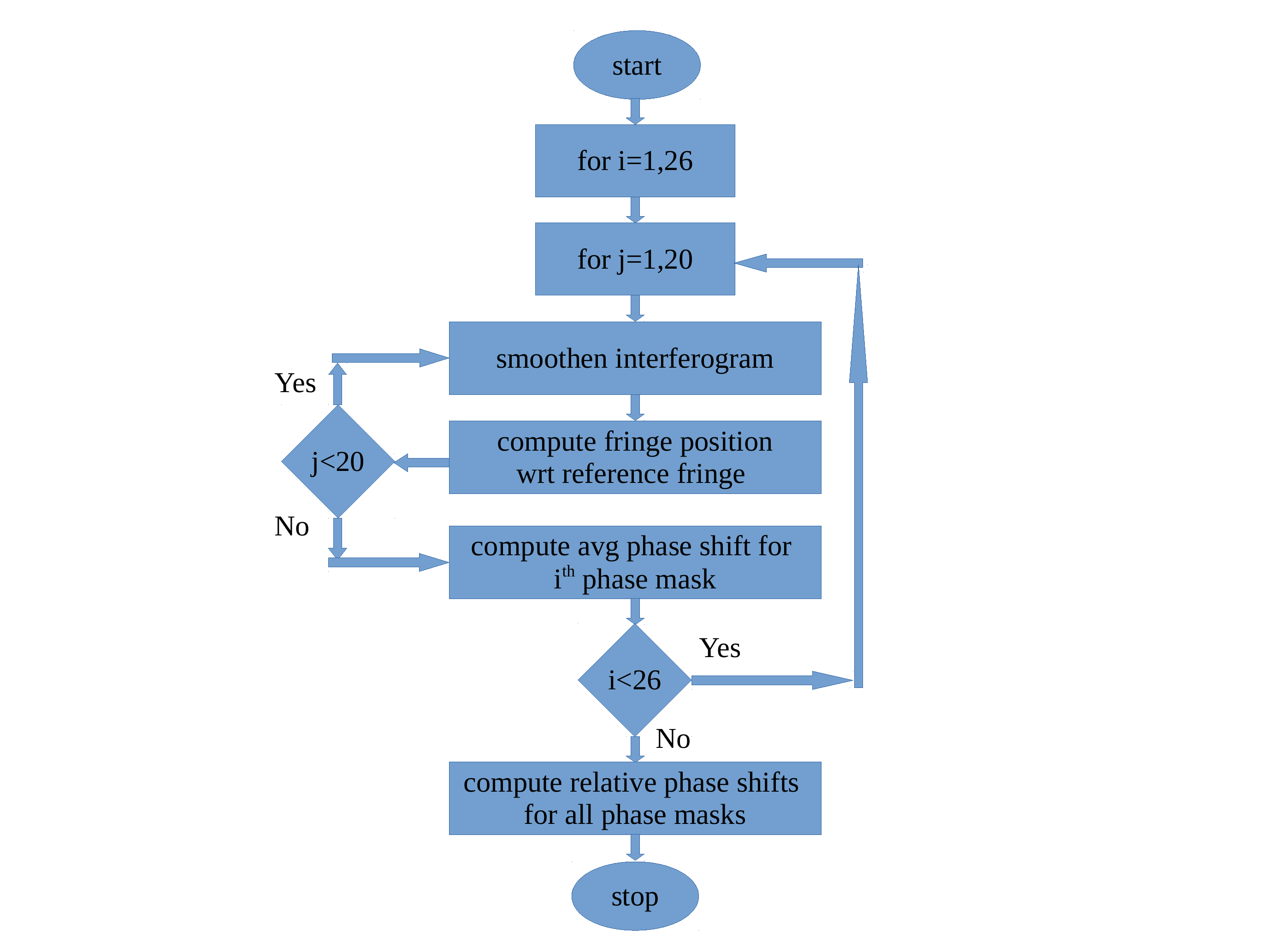}
\caption{Flowchart of the semi-automated pipeline depicting main steps involved in analyzing all experimental interferograms and obtaining the phase calibration curve of the SLM.}
\label{Fig. 2}
\end{figure}

\begin{figure}[ht!]
\centering\includegraphics[width=\linewidth]{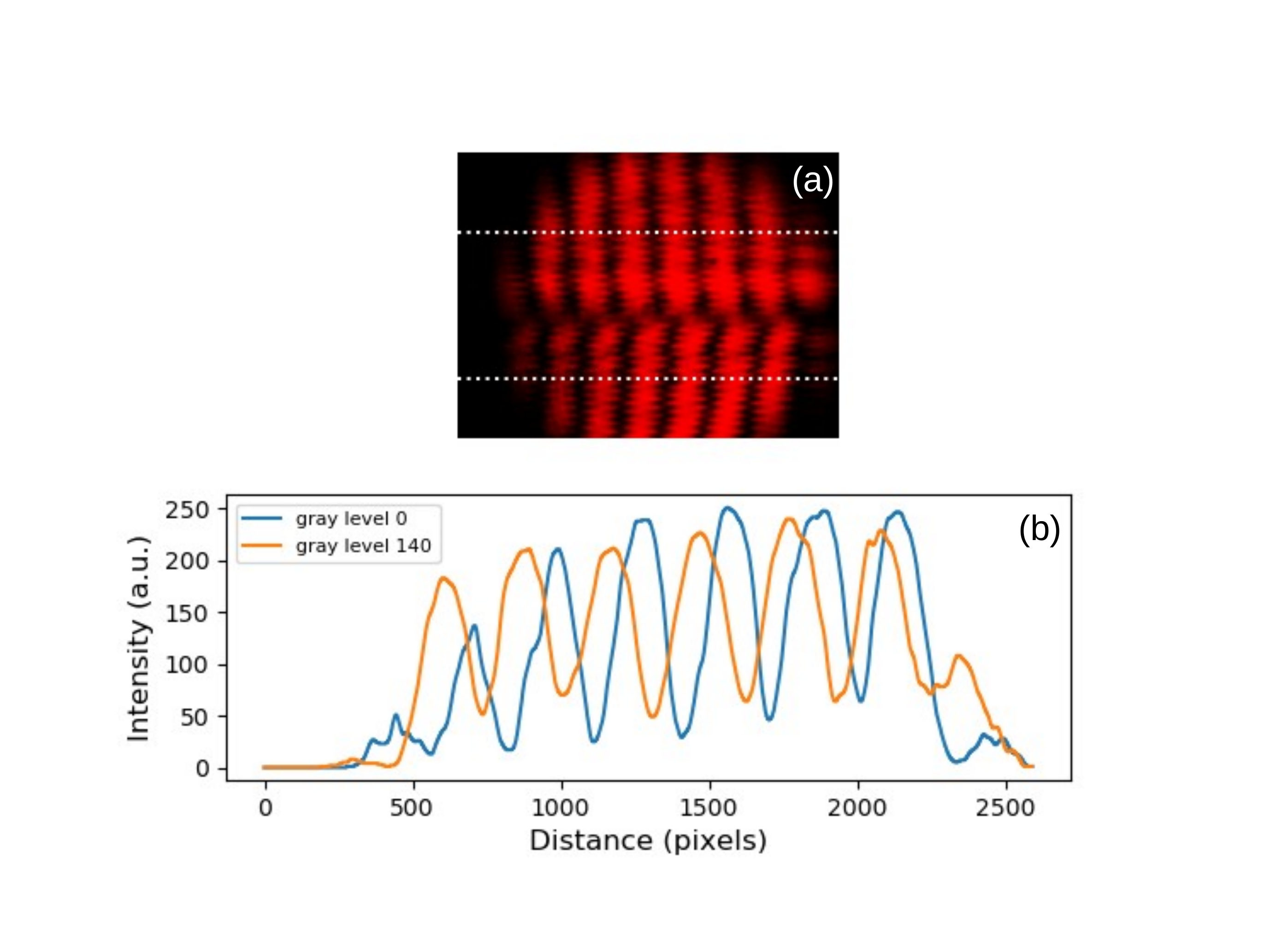}
\caption{(a) Locations on interferogram (shown by white dotted lines) selected for plotting profiles from top half and bottom half of interferogram. (b) Fringes showing relative phase shifts between reference gray level of 0 and desired gray level of 140.}
\label{Fig. 3}
\end{figure}

In order to measure phase shifts due to each addressed binary phase mask, we develop a semi-automated pipeline in python to analyze all 520 experimental interferograms and obtain the phase calibration curve of our SLM. The main steps of the semi-automated pipeline are shown in the flowchart in Fig. \ref{Fig. 2}. The first step in the analysis is to smoothen each interferogram and then make an appropriate 1D cut in the interferogram at two chosen locations, viz. one in the top half and the other in the bottom half of the interferogram (Fig. \ref{Fig. 3}). These locations of 1D cuts are frozen for all interferograms. In the subsequent step, a peak detection algorithm based on fringe intensity detection registers the location of peaks in the interferogram and computes the relative phase shift between the upper half (variable gray level) and the lower half (reference gray level) of the fringe pattern for each interferogram. We repeat this procedure for all 20 interferograms for a given binary phase mask and obtain the corresponding averaged relative phase shift. We then iterate this procedure over remaining interferograms and finally obtain the averaged relative phase shifts for all 26 gray level steps. This process of phase calibration is time-consuming and cumbersome and we suggest an improved method of phase calibration using phase masks that we  design indigenously. We construct the phase mask in the form of three different gray level stripes embedded on reference gray level of zero as shown in Fig. \ref{Fig. 7}(a). The width of each horizontal gray level stripe is 100 pixels while the separation between consecutive stripes is 25 pixels. The optimum choice of width of stripes include measurable intensity from different gray level stripes so as to obtain good contrast in the recorded interferogram, while the separation between stripes is chosen such that it is sufficient enough to avoid inter-pixel crosstalk. We record interferograms using the same experimental setup as used earlier for phase calibration.

\begin{figure}[ht!]
\centering\includegraphics[width=\linewidth]{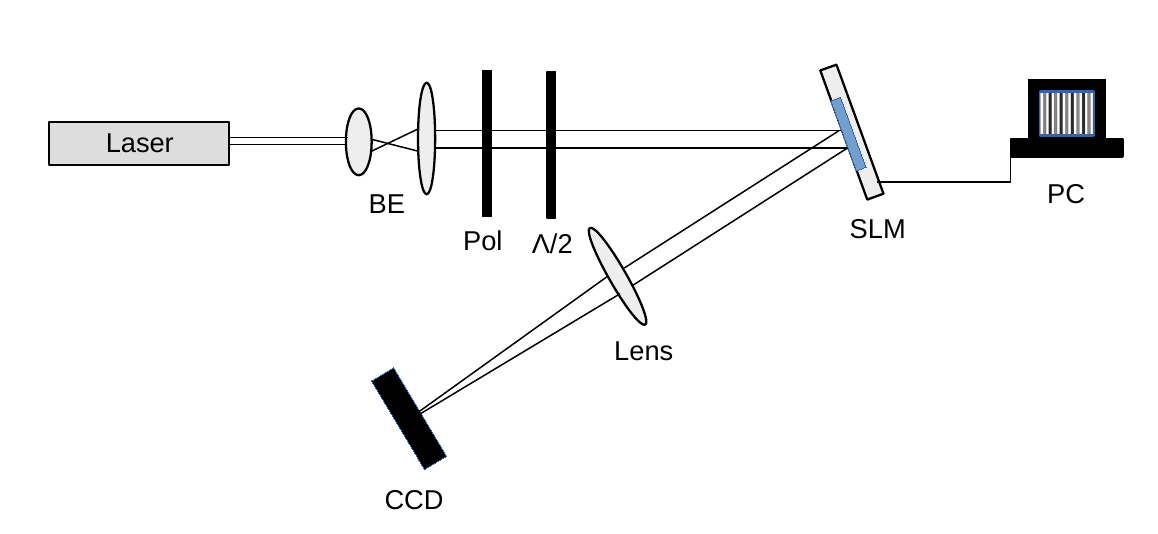}
\caption{Experimental setup for generating an array of uniform spots using IFTA.}
\label{Fig. 4}
\end{figure}

We now move on to the second segment of our work where our motive is to generate an array of uniform spots in 1D and 2D using an iterative fourier transform algorithm (IFTA), and investigate the efficacy of this algorithm experimentally. We develop an IFTA code based on a recent novel prescription \cite{gu2012generation}. We generate 1D optimised gratings using the iterative algorithm and study the convergence characteristics of the algorithm. We give random phases as the initial input to the algorithm. The phases of all slits in the optimized gratings are measured relative to that of the first slit whose phase is frozen at zero. For a test case to generate 1x5 array of uniform spots, we find that the optimum phase solutions obtained numerically converge in about 10 iterations. The experimental setup used to generate an array of spots in 1D/2D using optimised phase masks is shown in Fig. \ref{Fig. 4}. We use a 671 nm laser beam having diameter of about 2 mm which is expanded by a beam expander so that it illuminates the SLM display. The reflected beam from the SLM is fourier transformed by a 40 cm convex lens and is imaged on a CCD camera. The input power of the laser beam incident on the SLM is about 4.55 mW while the reflected beam for the SLM functioning akin to a mirror is about 3 mW. Hence, we compute the optical efficiency of our reflective SLM to be around $66\%$ which is concurrent with the vendor's specifications.

\begin{figure}[ht!]
\centering\includegraphics[width=\linewidth]{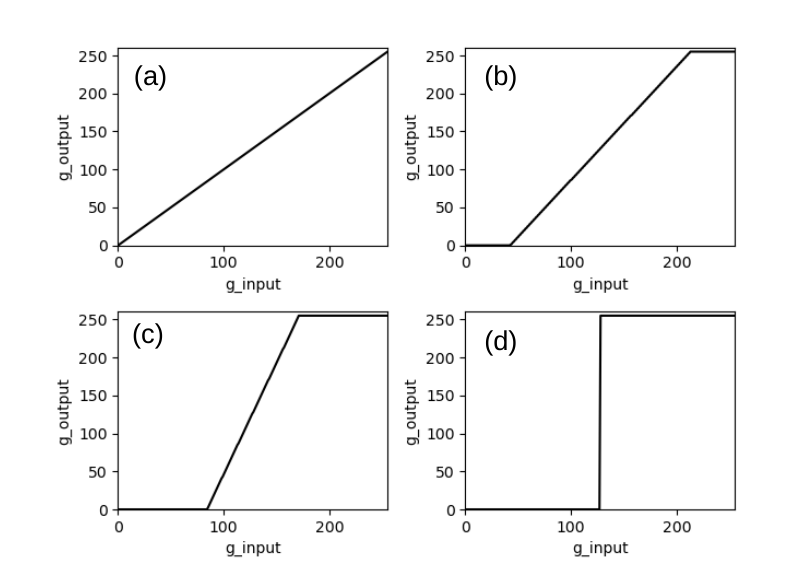}
\caption{Look-up table of an SLM showing different contrasts (a) C=1 (b) C=1.49 (c) C=2.93 (d) C=255.}
\label{Fig. 5}
\end{figure}

The final segment of our work deals with exploring methods to improve the diffraction efficiency of a phase limited SLM. It has been shown \cite{moreno2004modulation, bowman2011optimisation} that there is often a mismatch between the intended gray level and the actual gray level displayed on the SLM which diminishes its diffraction efficiency. A mitigation technique based on global linear corrections in the look-up table (LUT) has been suggested \cite{bowman2011optimisation} to improve the diffraction efficiency of SLMs which exhibit limited phase depth. A given LUT is quantified in terms of the slope of the linear function which maps the input gray level to the output gray level and is referred to as the contrast (C). The ideal LUT is shown in Fig. \ref{Fig. 5}(a) which has a contrast of unity. The method involves making incremental changes in the contrast of the LUT and measuring the corresponding first order diffraction efficiency obtained using a blazed grating. We implement this method based on global linear corrections in the LUT by modified phase encoding of the blazed grating (having periodicty of 16 pixels) displayed on the SLM. To begin with, we measure the power in the first diffraction order using the blazed grating (having C=1) displayed on the SLM. In the subsequent steps, we modify the phase encoding of the blazed grating by using the updated output gray level information computed using the desired LUT mapping function (having variable contrast) to obtain the modified blazed grating. The LUT, having a maximum contrast of 255, maps a blazed grating to a binary grating. We iterate this process using LUTs having different contrasts (some modified LUTs are shown in Fig. \ref{Fig. 5}) and measure the resultant first order diffraction efficiency. 

\section{Results}

\begin{figure}[ht!]
\centering\includegraphics[width=0.6\linewidth]{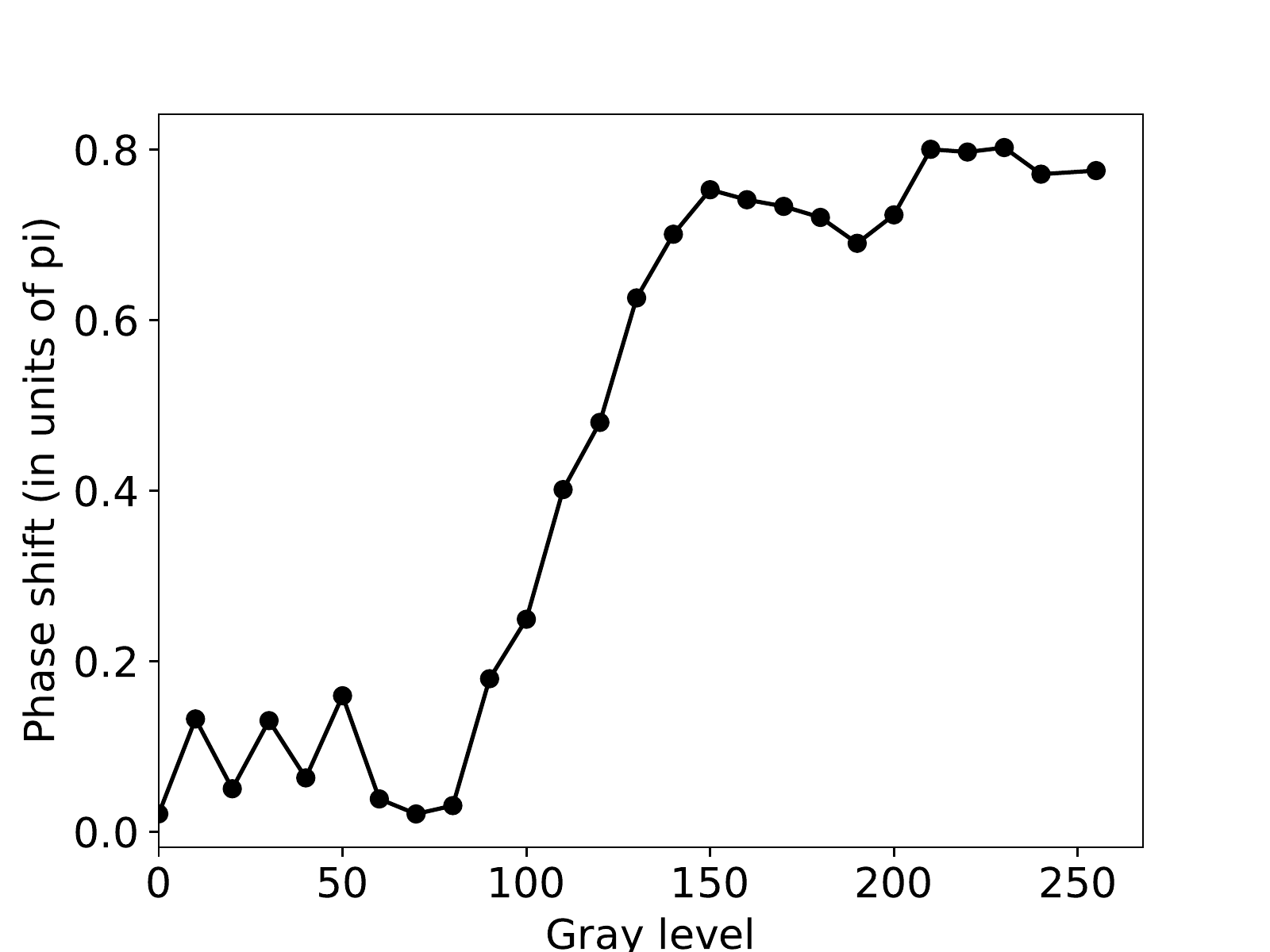}
\caption{Phase calibration curve of reflective SLM obtained using interferometric measurements at 671 nm.}
\label{Fig. 6}
\end{figure}

\begin{figure}[ht!]
\centering\includegraphics[width=\linewidth]{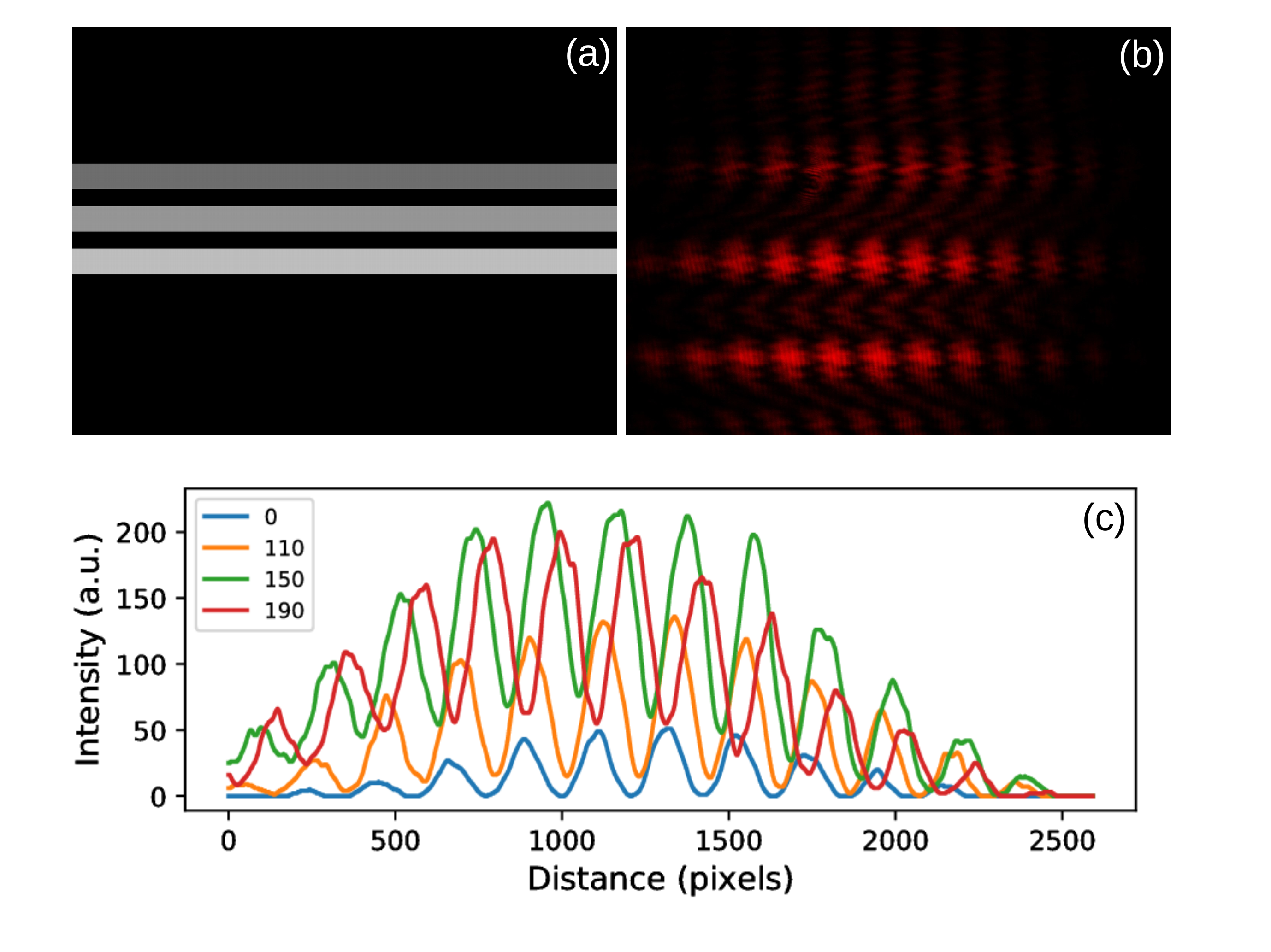}
\caption{(a) The new phase mask used in our experiment showing three different gray levels (top to bottom stripes: 110, 150 and 190) impressed on reference gray level of 0. (b) Interferogram exhibiting simultaneous phase shifts obtained using our phase mask. (c) Fringes showing simultaneous differential phase shifts with respect to the reference fringe.}
\label{Fig. 7}
\end{figure}

We obtain the phase calibration curve of our reflective SLM at 671 nm (Fig. \ref{Fig. 6}). The phase response exhibited by the SLM is non-linear while the linear regime of operation of the SLM lies between gray-value 80 to 150. We also observe that the maximum phase throw of our SLM at 671 nm is limited to about $0.8 \pi$ which is less than about $1\pi$ phase sweep according to vendor's specifications at visible wavelengths. Therefore, it is necessary to calibrate the SLM in laboratory environment conditions before using it in experiments as the phase response of an SLM is dependent on ambient conditions such as temperature, and since the maximum phase depth in general diminishes with increasing wavelength. Fig. \ref{Fig. 7}(b) shows the interferogram obtained using our novel phase mask which exhibits a set of four fringes obtained using a SLM based Michelson interferometer. Our modified phase mask enables us to simultaneously measure three differential phase shifts at a time (Fig. \ref{Fig. 7}(c)) which corresponds to three different gray levels displayed on the SLM. This further empowers us to reduce the number of steps involved in the phase calibration by one-third, which consequently enables faster phase calibration (by threefold) of the SLM.

\begin{figure}[ht!]
\centering\includegraphics[width=0.9\linewidth]{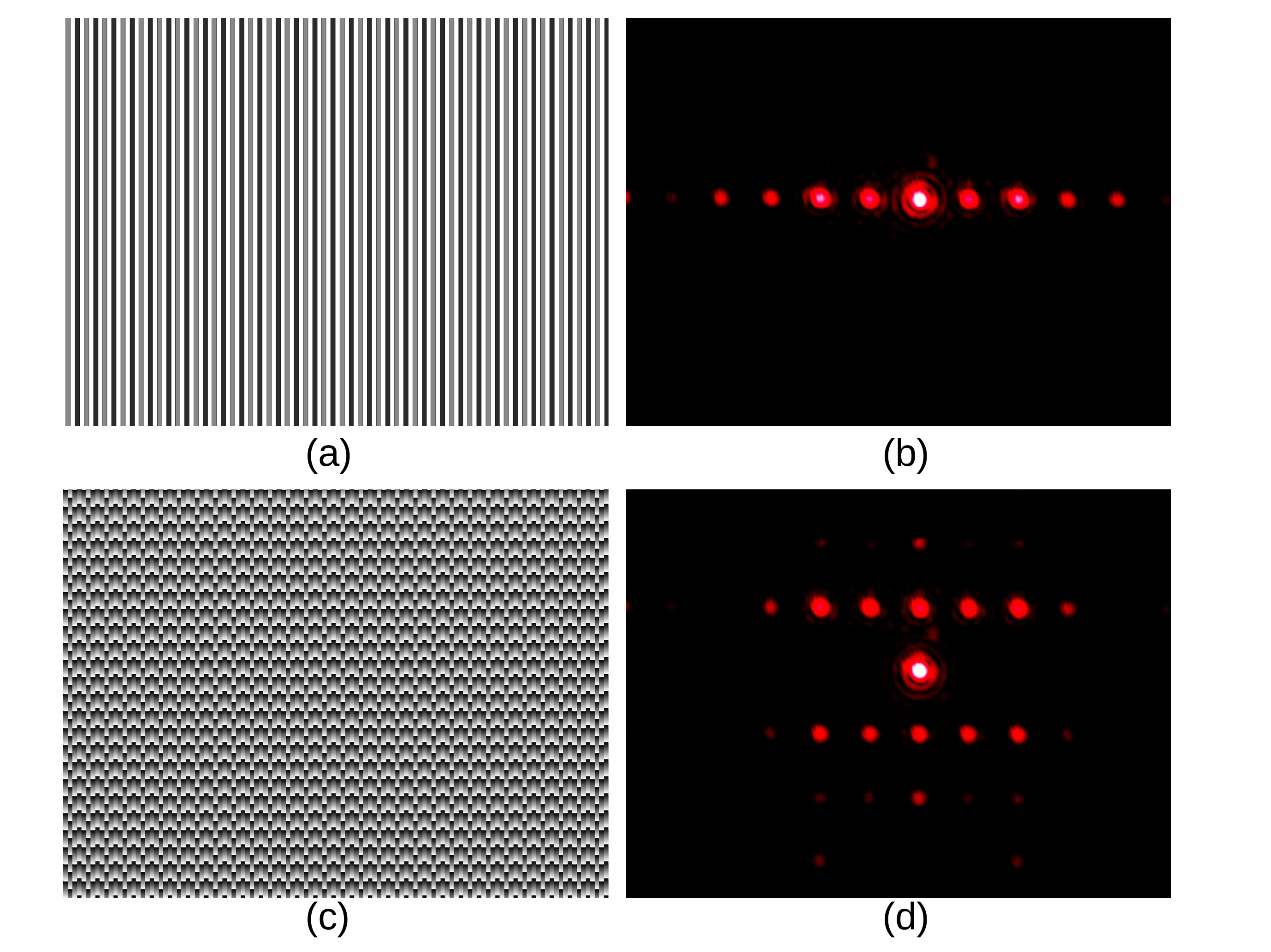}
\caption{(a) Optimised 1D phase mask obtained using iterative fourier transform algorithm. (b) 1x5 array of bright spots centered on zero-order beam generated using optimised phase mask. (c) Optimised phase mask
superimposed with a prism phase. (d) 1x5 array of uniform bright spots shifted along the vertical direction generated using modified phase mask.}
\label{Fig. 8}
\end{figure}

\begin{figure}[ht!]
\centering\includegraphics[width=\linewidth]{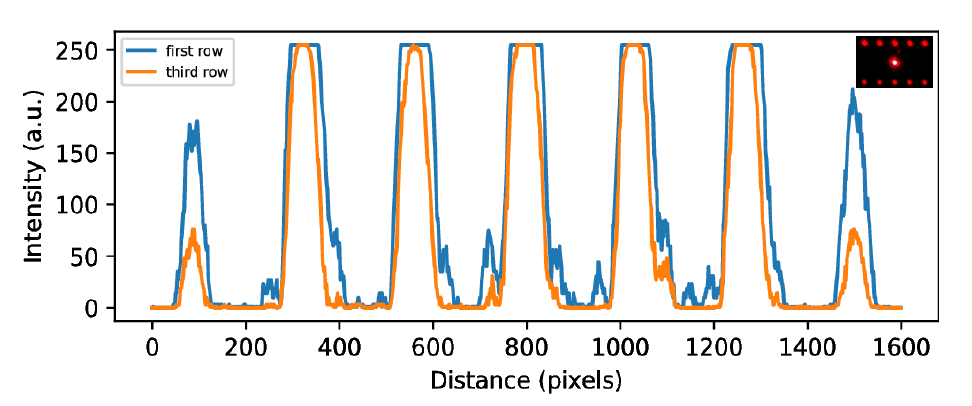}
\caption{Profile plot showing 1x5 array of uniform spots generated using iterative fourier transform algorithm. The image of the shifted array of uniform spots is shown in the inset.}
\label{Fig. 9}
\end{figure}

\begin{figure}[ht!]
\centering\includegraphics[width=\linewidth]{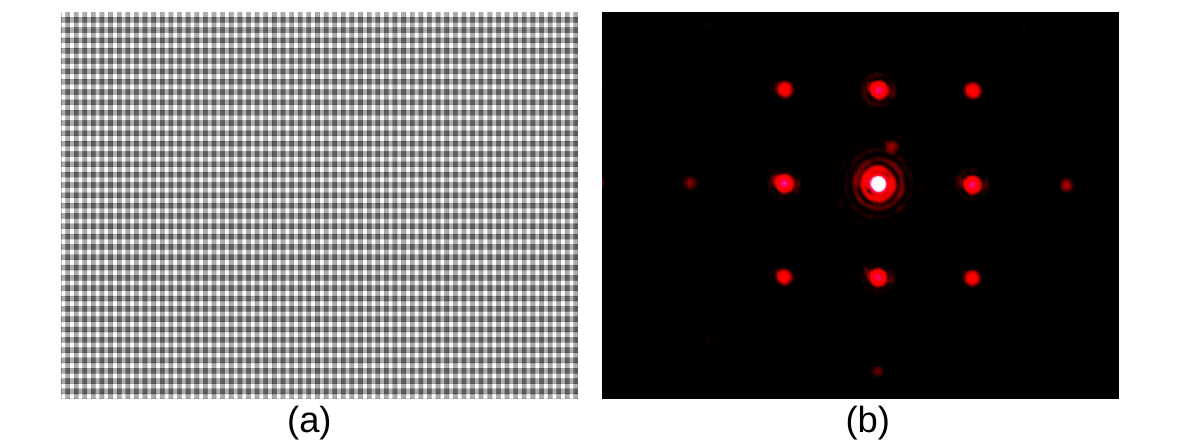}
\caption{(a) Optimised 2D phase mask obtained using iterative fourier transform algorithm. (b)  3x3 array of bright spots centered on zero-order beam generated using optimised phase mask.}
\label{Fig. 10}
\end{figure}

We experimentally obtain a 1x5 array of uniform spots centered on the zero-order beam using our optimised phase mask generated using IFTA (Fig. \ref{Fig. 8}(a) and (b)). We superimpose an additional prism phase on this optimised phase mask and obtain a uniform array of spots shifted vertically away from the zero-order beam (Fig. \ref{Fig. 8}(c) and (d)). Fig. \ref{Fig. 9} shows a 1D cross-section of the array of uniform spots shifted along the vertical direction (shown in Fig. \ref{Fig. 8}), and we find a total of five peaks which have almost equal intensity with parasitic spots having relatively less power compared to the desired diffracted spots. This unequivocally brings out the qualitative high uniformity of the array of spots generated using IFTA computations. We measure the actual power in each spot using a power meter and compute the uniformity ($u=\frac{I_{max}-I_{min}}{I_{max}+I_{min}}$) to be about $90\%$. We obtain diffraction efficiency of about $44\%$ using our optimised phase masks which is limited by the reduced phase throw of our SLM ($\phi_{max} < 2\pi$). We verify this manifestation using the experimentally measured maximum phase depth of our SLM ($\phi_{max}\sim 0.8\pi$) obtained from phase calibration at 671 nm. The resultant efficiency due to these optimised phase masks will certainly improve for SLMs having maximum phase depth reaching at least $2\pi$. We generate optimised phase masks in 2D by orthogonal superposition of 1D optimised phases obtained using IFTA. Fig. \ref{Fig. 10} shows a 3x3 array of spots generated using this iterative method.

\begin{figure}[ht!]
\centering\includegraphics[width=0.6\linewidth]{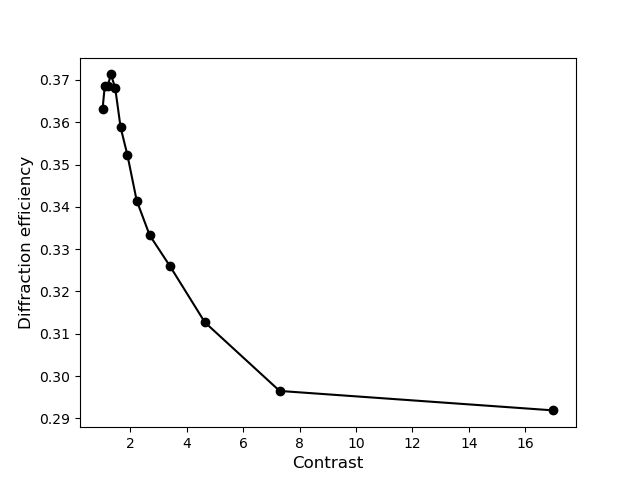}
\caption{Diffraction efficiency using global linear corrections in the look-up table of the SLM.}
\label{Fig. 11}
\end{figure}

Fig. \ref{Fig. 11} shows the diffraction efficiency obtained for our reflective SLM using global linear corrections in the LUT. We obtain the best diffraction efficiency of $37.1\%$ for a contrast of $1.3$. We observe that the efficiency decreases with further increase in contrast and starts saturates at a lower efficiency after C=8. The efficiency obtained using a binary grating is about $29.2\%$ and blazed gratings incorporating global linear corrections in the LUT improve the efficiency by about $27\%$. However, we find that the diffraction efficiency obtained using optimised phase masks generated using IFTA is about $44\%$ which is an improvement by around $19\%$ over the efficiency obtained using global linear corrections in the LUT. This illustrates the subtle uses of iterative algorithms to enhance diffraction efficiency of phase-limited SLMs which is a less explored area of research. 

\clearpage

\section{Conclusions}
We develop an improved method of interferometric phase calibration of a reflective SLM using novel phase masks which is threefold faster than conventional phase calibration methods. To the best of our knowledge, our work is the first attempt directed towards enabling rapid phase calibration of an SLM and can easily be implemented in settings which often require faster phase calibration of an SLM. We also develop a semi-automated pipeline in python for analysing interferograms and obtain the phase calibration curve at 671 nm. In the process, we ascertain the maximum phase throw of our SLM and find it to be phase-limited ($\phi_{max} \sim 0.8\pi$). Besides, our method, for the first time, exemplifies simultaneous differential tuning of an SLM based interferometer which can have diverse uses in spectroscopic applications. In addition, it can assuage limited temporal scanning rates of an interferometer by facilitating collection of simultaneous multi-wavelength spectral information of an object under study. This can enable us to better investigate phenomena which evolve faster than the scanning rate of the interferometer. Over and above, the biggest advantage of using SLMs for interferometric applications is that it does not involve any moving mechanical parts which can be rather useful for applications such as in astronomy where any wear and tear in moving parts can render the instrument useless. Thus, we implement an IFTA algorithm and test the experimental efficacy of this algorithm by generating an array of spots in 1D and 2D. We find that the iterative algorithm performs well in terms of fast convergence and uniformity ($90\%$) but gives moderate efficiency (about $44\%$) which is due to the limited phase depth of our SLM. We further validate this observation using the experimentally determined maximum phase throw of our SLM at 671 nm. Thus, this iterative algorithm can be useful for close to real-time applications such as in adaptive wavefront sensing, beam multiplexing and holographic optical tweezers. In addition, rapid convergence of the iterative algorithm can be of significant use in settings possessing modest computational budget. We implement global linear corrections in the LUT to improve the diffraction efficiency of our phase-limited SLM and find that optimised phase masks generated using iterative algorithm improve efficiency by about 19\% over global linear corrections. Thus, we also illustrate relatively less known uses of iterative algorithms in improving diffraction efficiency of phase-limited SLMs, which, to the best of our knowledge is the first endeavour in this direction. In future work, we would like to use our findings to generate working holographic optical tweezers and optimized vector beams for optical trapping. We also plan to extend our research to wavefront correction for image reconstruction to facilitate improved imaging which may open up useful applications in astronomy with SLMs. 

\section*{Acknowledgments}
This work was supported by the Center of Excellence in Space Sciences India (CESSI) and the Indian Institute of Science Education and Research, Kolkata, funded by the Ministry of Human Resource Development, Govt. of India. A.D.C. acknowledges support from the INSPIRE fellowship of the DST, Govt. of India.

\end{document}